\documentclass[prd,twocolumns,showpacs,groupedaddress,superscriptaddress,amsmath,amssymb,reprint, nofootinbib, nothanksinbib]{revtex4-2}

\usepackage{graphicx}
\usepackage{dcolumn}
\usepackage{bm}
\usepackage{hyperref}
\usepackage{xcolor}

\def\u{\tilde{u}}
\def\s{\tilde{s}}

\def\P{\mathcal{P}}

\begin{document}

\title{Updated E141 constraints on a long-lived $X_{17}$ vector boson}

\author{A. Celentano}
\affiliation{
Universit\`a degli Studi di Genova, Dipartimento di Fisica, 16146 Genova, Italia
}
\affiliation{
INFN, Sezione di Genova, 16146 Genova, Italia
}
\author{A. Marini}
\affiliation{
INFN, Sezione di Genova, 16146 Genova, Italia
}
\author{L. Marsicano}
\thanks{Corresponding author}\email{luca.marsicano@ge.infn.it}
\affiliation{
INFN, Sezione di Genova, 16146 Genova, Italia
}

\date{\today}

\begin{abstract}

Since its first observation by the ATOMKI experiment in 2018, the $\rm ^*Be$ anomaly  has attracted considerable interest within the dark sector community as it may indicate the existence of a new fundamental particle with a mass of about 16.9 MeV, the $X_{17}$. However, the minimal model describing $X_{17}$ as a new vector boson is severely constrained by null results from legacy beam-dump experiments. Among these, the E141 experiment at SLAC places stringent limits on the $X_{17}$ coupling to electrons $\varepsilon_e$ in the $5.1\times10^{-5}\lesssim \varepsilon_e\lesssim1.7\times10^{-4}$ region. This excludes the possibility of a long-lived $X_{17}$, potentially in contrast with the preliminary estimate of the particle lifetime recently reported by the ATOMKI collaboration. The E141 limits commonly adopted in the literature rely on reinterpretations of the original analysis under solid but  simplifying assumptions. While these studies provide a reliable estimate of the experiment's reach, they rely on approximations that were well justified when the boson mass was largely unconstrained. With the $X_{17}$ mass now confined to a narrow region by recent experimental observations, a more refined treatment of the E141 sensitivity becomes necessary. In this work, we revisit the E141 exclusion limits in the $X_{17}$ scenario by performing a dedicated reanalysis that incorporates a more accurate treatment of the experimental setup and signal prediction. We quantify the impact of these refinements on the excluded parameter space and discuss their implications for the compatibility between the E141 constraints and the vector boson interpretation of the ATOMKI anomalies.
\end{abstract}

\maketitle


\section{Introduction}

The first hint for the existence of a new, light boson coupled to electrons and positrons, the so-called $X_{17}$ particle, was reported almost ten years ago by Krasznahorkay and collaborators~\cite{Krasznahorkay:2015iga} from the analysis of experimental data collected at the ATOMKI laboratory by bombarding a thin $^7$Li foil with a low-energy proton beam. The ATOMKI experiment measured the angular correlation of $e^+e^-$ pairs produced by the decay of $^8$Be excited states through the internal pair creation (IPC) process - by varying the beam energy, the experiment selectively populated the 17.6~MeV and 18.15~MeV $^8$Be $1^+$ states. A discrepancy of approximately $5\sigma$ between the measured relative $e^+e^-$ angle distribution and the prediction from standard nuclear-physics models, for the angular range $\Theta_{e^+e^-}\approx 140^\circ$, was observed. This anomaly, recently confirmed by an independent measurement at the VNU University of Science (HUS) laboratory~\cite{Anh:2024req}, was immediately linked to the possible existence of a new light particle, with mass approximately 16.7~MeV, coupled to quarks and leptons, that is produced almost-at-rest in the decay of the $^8$Be excited state and then decays to an $e^+e^-$ pair~\cite{Feng:2016jff,Feng:2016ysn}. Later, the Hungarian group repeated the IPC angular distribution study exploiting the $^3\mathrm{H}(p,e^+e^-)^4\mathrm{He}$~\cite{Krasznahorkay:2021joi} and the $^{11} \mathrm{B}(p,e^+e^-)^{12}\mathrm{C}$~\cite{Krasznahorkay:2022pxs} reactions. All measurements showed a large discrepancy in the measured $\Theta_{e^+e^-}$ distribution with respect to theory predictions. Interestingly, in agreement with the particle-physics interpretation of this effect, the anomaly appeared at different values of the $e^+e^-$ opening angle, all consistent with a mass value of roughly 17 MeV~\cite{Krasznahorky:2024adr}. 

Immediately after its first formulation, the $X_{17}$ hypothesis motivated many new experiments aiming at further investigating the existence of the new light state, either by exploiting again the $^7\mathrm{Li}(p,e^+e^-)^8\mathrm{Be}$ reaction with a different experimental apparatus~\cite{MEGII:2024urz,Gustavino:2024wgb,Servin:2024kxt,NewJEDI:2023ltw,Anh:2024req}, or by testing the $X_{17}$ production through its coupling to electrons and positrons~\cite{PADME:2025dla,NA64:2019auh,Dutta:2023ifr}. In parallel, it triggered in the community a large number of theoretical studies devoted to its interpretation, either as a possible hint for the existence of new physics~\cite{Feng:2020mbt,Zhang:2020ukq,Kozaczuk:2016nma,Ellwanger:2016wfe,Fieg:2026zkg}, or as the indication of Standard Model (SM) effects not properly captured by current nuclear models~\cite{Zhang:2017zap,Chen:2020arr,Kubarovsky:2022zxm,Aleksejevs:2021zjw} (see also~\cite{Barducci:2022lqd} and ~\cite{Krasznahorky:2024adr} for a recent review). Many phenomenological studies have also been carried out to check the impact of such a hypothesis on other particle-physics fields, including neutrino physics~\cite{Rathsman:2026smv,Denton:2023gat, Mustamin:2026zht}, muon physics~\cite{Capolupo:2025lyw}, light mesons physics~\cite{Hostert:2023tkg,Hostert:2026xul}, and QCD axion studies~\cite{Alves:2020xhf,Liu:2021wap}.

Despite these efforts, today a unanimous consensus within the community regarding the origin and the explanation of the $X_{17}$ anomaly has not yet been reached. If the new particle interpretation is assumed to be true, the mass of the $X_{17}$ can be robustly extracted from a global fit based on all currently-available data, including the measured $\Theta_{e^+e^-}$ distributions reported by the ATOMKI experiments, the recent findings from PADME, and the null observation from MEG-II: the result predicts a value of $(16.88\pm 0.05)$~MeV~\cite{Arias-Aragon:2025wdt}, in continuity with a preliminary result of $(16.85\pm 0.04)$~MeV based only on the nuclear physics experiments~\cite{PhysRevD.108.015009}. 
For what concerns the new particle coupling to quarks, a first estimate has been reported by Feng \textit{et al.}~\cite{Feng:2016ysn}, who exploited the measured relative branching ratio \text{$B(^8\mathrm{Be}^*\rightarrow {}^8\mathrm{Be}X_{17})/B(^8\mathrm{Be}^*\rightarrow {}^8\mathrm{Be}\,\gamma)$} under the spin-parity assumption $J^P=1^-$. The $J^P=1^+$ scenario was later scrutinized by Mommers and Vanderhaeghen~\cite{Mommers:2024qzy}, who adopted a similar approach exploiting data from the decay of excited $^{12}$C states. 

The coupling to electrons is constrained by the qualitative estimate in Ref.~\cite{Feng:2016ysn}, $\varepsilon_e \gtrsim 1.3 \times 10^{-5}$, based on the requirement that the $X_{17}$ decay length was short enough to match the apparatus acceptance in the original ATOMKI measurement; more recently, a further constraint of $\varepsilon_e \gtrsim 2.6 \times 10^{-5}$ was obtained by a preliminary measurement of the $X_{17}$ lifetime~\cite{x17longlived,x17longlived1}.
The allowed $\varepsilon_e$ range has been explored by NA64~\cite{NA64:2019auh}, PADME~\cite{PADME:2025dla} and KLOE~\cite{Anastasi:2015qla}; recent results from FASER can also constrain the parameter space, if a specific combination of the $X_{17}$ couplings to quarks is assumed~\cite{Kling:2025udr}. Further limits on  $\varepsilon_e$ have been set by a re-analysis of past electron beam-dump experiments by Andreas and collaborators~\cite{Andreas:2012mt}. Among the latter, results from the SLAC E141 9-GeV electron beam-dump~\cite{Riordan:1987aw} currently exclude the coupling region between $\varepsilon_e \approx 5\times 10^{-5}$ and $\varepsilon_e \approx 2\times 10^{-4}$. It is worth noting, however, that the preferred $X_{17}$ mass interval lies close to the boundary of the E141 excluding region, where the assumptions and approximations adopted in the re-analysis -- while well suited to the scope of the original work -- may have a non-negligible impact on the resulting sensitivity. Indeed, while the re-analysis presented in~\cite{Andreas:2012mt} provided a valuable estimate of the E141 sensitivity to a generic light vector boson, the precise definition of the $X_{17}$ mass interval warrants a more detailed assessment of the E141 constraints.

In this work we present an improved re-analysis of the original E141 dataset, including for the first time kinematic effects previously not taken into account, such as the impinging beam divergence and the electrons track-length angular distribution in the beam dump. Our result shows that the E141 original dataset does not permit deriving any exclusion limit for the existence of new light boson with mass $\simeq$ 17 MeV. 

The paper is organized as follows. In Section~\ref{sec:pheno} we briefly summarize the phenomenology associated with $X_{17}$ production in thick-target electron-beam experiments. Then, in Section~\ref{sec:E141} we present our re-analysis of the original E141 dataset, first providing a complete description of the E141 experimental setup, and then describing in detail the procedure that we followed to compute the $X_{17}$ production yield within the dump and the experimental acceptance.

\section{\label{sec:pheno} $X_{17}$ production in thick-target electron-beam experiments}

In this work, we assume for simplicity that the $X_{17}$ is a vector boson. The minimal effective lagrangian density describing the current $X_{17}$ phenomenology reads: 

\begin{equation}\label{eq:theory}
\mathcal{L} \supset -\frac{1}{4}X_{\mu\nu}X^{\mu\nu}+\frac{m^2_{X}}{2}X_{\mu}{X^\mu}+\sum_{f=e,\nu,u,d}e\varepsilon_f X_{\mu} \bar{f}\gamma^\mu f\;\;,
\end{equation}
where $X_{\mu}$ is the $X_{17}$ field, $X_{\mu\nu}\equiv \partial_\mu X_\nu - \partial_\nu X_\mu$, $m_X$ is the new boson mass, and $e\varepsilon_f$ is the generic coupling for the interaction between the $X_{17}$ and the Standard Model fermion $f$. Following the arguments of Ref.~\cite{Feng:2016jff} and Ref.~\cite{Denton:2023gat}, here we assume a negligible coupling to neutrinos, $\varepsilon_\nu\approx0$, so that, once produced, the $X_{17}$ state will always decay to an $e^+e^-$ pair with intrinsic width given by:
\begin{equation}
\Gamma = \frac{1}{3}\alpha \varepsilon^2_e m_X \left(1+\frac{2m_e^2}{m_{X}^2}\right)
\sqrt{1-\frac{4m^2_e}{m_{X}^2}} \, .
\end{equation}

For a multi-GeV electron beam impinging on a thin target, the dominant $X_{17}$ production process is the radiative emission in the electromagnetic field of a target nucleus, $e^-(p) N(P_i) \rightarrow e^-(p') N(P_f) X_{17}(k)$ (quantities in parenthesis denote each particle four momentum). This process is characterized by a forward-peaked kinematics, with the $X_{17}$ emitted at very small angle and taking most of the primary beam energy~\cite{Bjorken:2009mm}. The exact-tree-level (ETL) expression for the differential cross section $\frac{d\sigma}{dx \, dcos(\theta)}$, expressed as a function of the $X_{17}$ energy fraction $x=E_X/E$ and angle $\theta$, reads:

\begin{equation}\label{eq:ETL}
\frac{d\sigma}{dx\,dcos(\theta)}= \int_{t_{\rm min}}^{t_{\rm max}} dt\,
\frac{\alpha^{3}\varepsilon_e^2\,|\bm{k}|\,E}
     {8\,M^{2}\,|\bm{p}|\,|\bm{p}-\bm{k}|}
\,
\frac{\left|F(t)\right|^2}{t^{2}}
\,
\sum_{i=-2}^{1}I_i \,C_i \; \;,
\end{equation}
where $E$ is the initial-state electron energy, $M$ is the nucleus mass, $t=-(P_f-P_i)^2$ is the momentum transferred squared, and $F(t)$ is a product of the atomic and nuclear form factor, for which we adopt the parameterization reported in Ref.~\cite{Bjorken:2009mm}. In Eq.~\ref{eq:ETL}, $t_{min}$ and $t_{max}$ are the $x-$ and $\theta-$dependent momentum transferred kinematic limits, whose complete expressions, together with those for the $I_i$ and $C_i$ functions, are given in Appendix~\ref{appendix:MM}.
\begin{figure*}[t!]
    \centering
    \includegraphics[width=.75\textwidth]{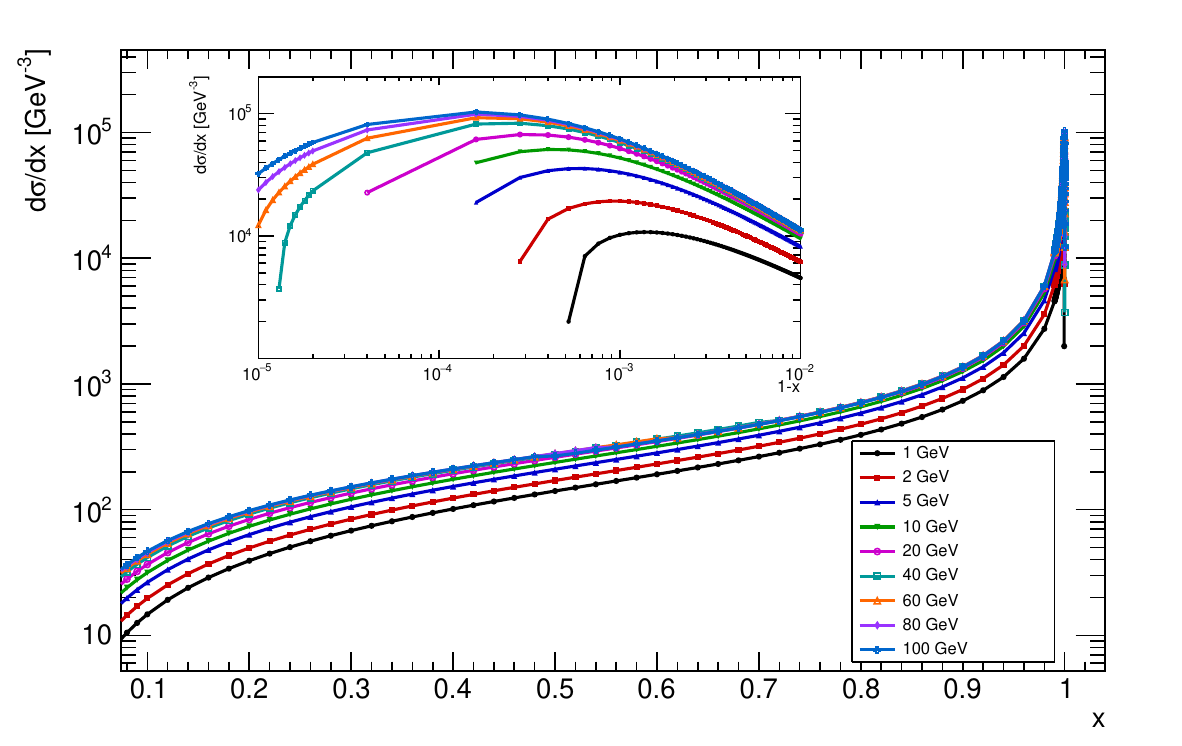}
    \caption{Differential cross section $d\sigma/dx$ for $X_{17}$ radiative production on lead ($\varepsilon_e=1)$, for different initial-state electron energies as reported in the legend. The inset shows a zoom of the $x\approx1$ region - for better visualization, here the cross section is plotted as a function of $1-x$.}
    \label{fig:ETL}
\end{figure*}

For given electron energy, $x$ ranges from $x_{min}=m_{X}/E$ to the maximum value $x_{max}$ given by:
\begin{equation}
x_{ max}
=
\frac{
(E+M)\Delta
+
|\bm{p}|\,
\sqrt{\Delta^2-4m_X^2s}
}
{2E_0\,s},
\label{eq:xmax}
\end{equation}
where $\Delta \equiv 2M(E-m_e)+m_X^2$ and $s\equiv M^2+m_e^2+2E\,M$. The differential cross section grows for $x$ varying from $x_{min}$ to $x_{peak}$ and then rapidly decreases for $x\rightarrow x_{max}$. For the multi-GeV energy range relevant for this study, we found that $x_{peak}$ is reasonably parametrized by the expression $x_{peak}\approx1-\delta$, where $\delta \approx \xi \frac{m_e}{m_X}$ with the factor $\xi$ being of the order of $10^{-2}$ - this scaling is in agreement with the results reported in Ref.~\cite{Bjorken:2009mm}. For illustration, Fig.~\ref{fig:ETL} shows the differential cross section $d\sigma/dx$ for $X_{17}$ production on a thin Pb target for various values of the electron energy $E$. For $x$ close to $x_{peak}$, the angular $X_{17}$ distribution is forward-peaked, with typical emission angle scaling as $\theta_{prod} \approx \frac{\sqrt{m_em_X}}{E_0}$~\cite{Izaguirre:2013uxa}. After production, a relativistic $X_{17}$ particle decays to an $e^+e^-$ pair, with typical opening angle $\Theta_{e^+e^-}\approx m_{X}/E_X$. The corresponding decay length scales as:
\begin{equation}\label{eq:lambda}
\lambda_D= \gamma c\tau \approx 30\mathrm{\,cm}\left(\frac{E_X}{10\mathrm{\,GeV}}\right)\left(\frac{10^{-4}}{\varepsilon_e}\right)^2\left(\frac{17\mathrm{\,MeV}}{m_X}\right)^2 \; \;,
\end{equation}
thus resulting to a detached-vertex signature for the $E\gtrsim10$~GeV range considered in this work.

The phenomenology of $X_{17}$ production in an electron-beam thick-target setup follows from the convolution of the differential cross section with the kinematic distribution of electrons within the electromagnetic shower developing in the target. The latter is described by the so-called ``track-length'' differential distribution $T(E_e,\cos\theta_e,\phi_e)$, defined such that 
$T(E_e,\cos\theta_e,\phi)\, dE_e\, d(\cos\theta_e)\, d\phi$ is the cumulative path length travelled inside the target by electrons whose energy, polar angle cosine, and azimuthal angle lie within the corresponding differential intervals. The differential $X_{17}$ production yield is given by the expression
\begin{equation}\label{eq:yield}
    \frac{dN_{X}}{dE_X}=N_{\mathrm{EOT}} \frac{N_{A}}{A}\rho\int_{m_X}^{E_{0}} dE_e\, T(E_e)\frac{d\sigma}{dE_X}(E_e,E_X) \; \;,
\end{equation}
where $A$ and $\rho$ are the target material atomic mass and density, $N_{\mathrm{EOT}}$ is the total number of electrons impinging on the target, $E_0$ is the impinging beam energy, $T(E_e)$ is the angle-integrated track-length, normalized to a single impinging electron, and $d\sigma/dE_X(E_e,E_X)$ is the $X_{17}$ production cross section, integrated over the emission angle and computed for initial-state electron energy $E_e$.
\begin{figure}[t]
    \centering
    \includegraphics[width=0.5\textwidth]{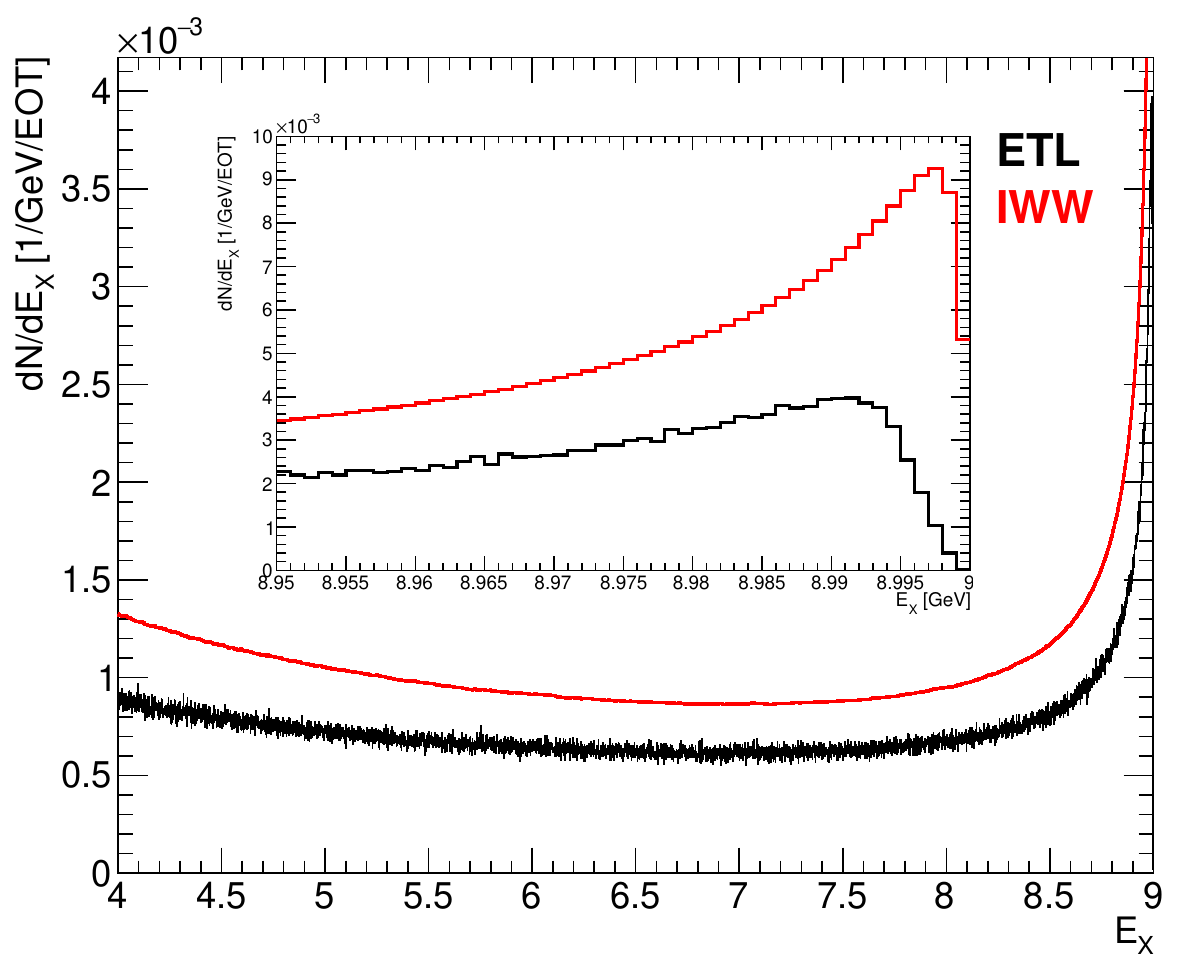}
    \caption{Energy spectrum of $X_{17}$ particles produced by a 9-GeV electron beam impinging on a thick $W$ target as obtained from Eq.~\ref{eq:yield} adopting for the differential cross section the ETL formula (black) and the Improved Weizs\"acker-Williams approximation (red). Both distributions are normalized to a single EOT, for $\varepsilon_e=1$.}
    \label{fig:dNdE}
\end{figure}
For illustration, the predicted $X_{17}$ energy spectrum for a 9 GeV electron beam impinging on a thick tungsten target, normalized to the total number of EOT, is shown in Fig.~\ref{fig:dNdE}, comparing the result from Eq.~\ref{eq:yield} obtained using the ETL formula with those from the Improved Weizs\"acker-Williams (IWW) approximation (see e.g. Eq. 5 in Ref.~\cite{Andreas:2012mt}). As a consequence of the pronounced peak at $x_{peak}\approx1$ characterizing the microscopic cross section, the $dN/dE_X$ shape directly reflects the electrons track-length energy distribution in the target - the rise at $E_X\simeq E_0$ is due to the contribution of the primary beam particle that can radiate a high-energy $X_{17}$ in the very first portion of the target. While the two curves indeed show a similar trend, the normalization of the IWW distribution is significantly larger than that of the ETL one - the IWW/ETL ratio smoothly varies from $\approx 1.5$ for $x=0.5$ to $\approx 1.4$ for $x=0.9$, and then rises up to 2 for $x \approx 1$ - this observation is in agreement with the findings of Ref.~\cite{Liu:2017htz}, where a comprehensive comparison between the two approaches was performed. 

The differential electrons and positrons yield resulting from $X_{17}$ decay events occurring in the region downstream the target extending from $L_{sh}$ (target thickness) to $L_{sh}+L_{dec}$ is then given by the expression\footnote{Eq.~\ref{eq:diffYield} neglects the depth of $X_{17}$ production within the dump, assuming it always occurs at the beginning of the target. This assumption is usually well justified for an experiment with a very thick target, in which the corresponding thickness $L_{sh}$ is much longer than the distance between the target front face and the production vertex. However, this is not the case for the E141 experiment - in the analysis presented in this work the effect has been properly accounted for, as discussed in the next Section.}:
\begin{equation}\label{eq:diffYield}
\frac{dN_p}{dE_p}=\int_{E_e}^{E_0} dE_X\frac{dN_X}{dE_X}e^{-L_{sh}/\lambda_D}
\,\left(1-e^{-L_{dec}/\lambda_D}\right) \, f(E_p;E_X) \; 
\end{equation}
where $f(E_p;,E_X)$ is the final state lepton energy $E_p$ PDF for given $X_{17}$ parent energy - for an isotropic decay, neglecting $m_e$, $f = \frac{1}{E_X}$, while in the fully transverse polarized scenario $f=\frac{3}{4E_X}\left(1+(2\frac{E_p}{E_X}-1)^2\right)$ for the same approximation. The total signal yield $N_s$ measured by a detector placed downstream of the beam-dump that is sensitive to either positrons or electrons from $X_{17}$ decay products reads
\begin{equation}\label{eq:totYield}\
N_s=\int_{E_p^{m}}^{E_p^{M}} dE_{p} \frac{dN_p}{dE_p}\; \; ,
\end{equation} 
where the integration limits $E^m_{p}$ and $E^M_{p}$ are set by the detector energy threshold and measurement range. We emphasize that, in order to properly account for the longitudinal acceptance of the decay region parametrized by the two exponential factors in Eq.~\ref{eq:diffYield}, a proper description of the high-energy portion of the $X_{17}$ spectrum is crucial, since this region contributes most to the overall signal yield. For example, for the values $L_{sh}=12.16~$cm and $L_{dec}=35~$m, representative of the E141 experimental setup as discussed in the next section, the probability to decay within the decay volume increases from $\approx 5\%$ at $E_X\approx E_0/2$ to $\approx 25\%$ at $E_X\approx E_0$ for $\varepsilon_e=1.7\times 10^{-4}$, representing the upper boundary for the current E141 exclusion limit -- for the lower boundary ($\varepsilon_e=5.1\times 10^{-5}$), instead, a less pronounced increase of about $20\%$ is observed.

The total signal yield should be further corrected to account for angular effects, that result in a reduction of the number of $X_{17}$ decay particles falling within the acceptance of the detector. Three main effects contribute to the overall angular spread of these, namely (i) the intrinsic angular spread of primary and secondary electrons within the dump due to multiple scattering, (ii) the $X_{17}$ emission angle $\theta_{prod}$ with respect to the radiating electron, and (iii) the $e^+e^-$ decay pair opening angle $\Theta_{e^+e^-}$. For the multi-GeV energy range considered in this work, the three contributions are all within the same order of magnitude -- for $E_0\approx 10$~GeV the three angles are of the order of a few mrad, while at $E_0\approx 100$~GeV a one-order-of-magnitude reduction is expected. Additionally, the impinging beam angular divergence should be accounted for. This calls for an accurate treatment of all these effects in the evaluation of the signal yield in a beam-dump experiment such as E141, in which, as discussed in the following, the detector angular acceptance was of about $1$ mrad.

\section{\label{sec:E141}Improved $X_{17}$ exclusion limits from the E141 experiment}
\subsection{The SLAC E141 experiment}

\begin{figure}[t]
    \centering
    \includegraphics[width=1.05\linewidth]{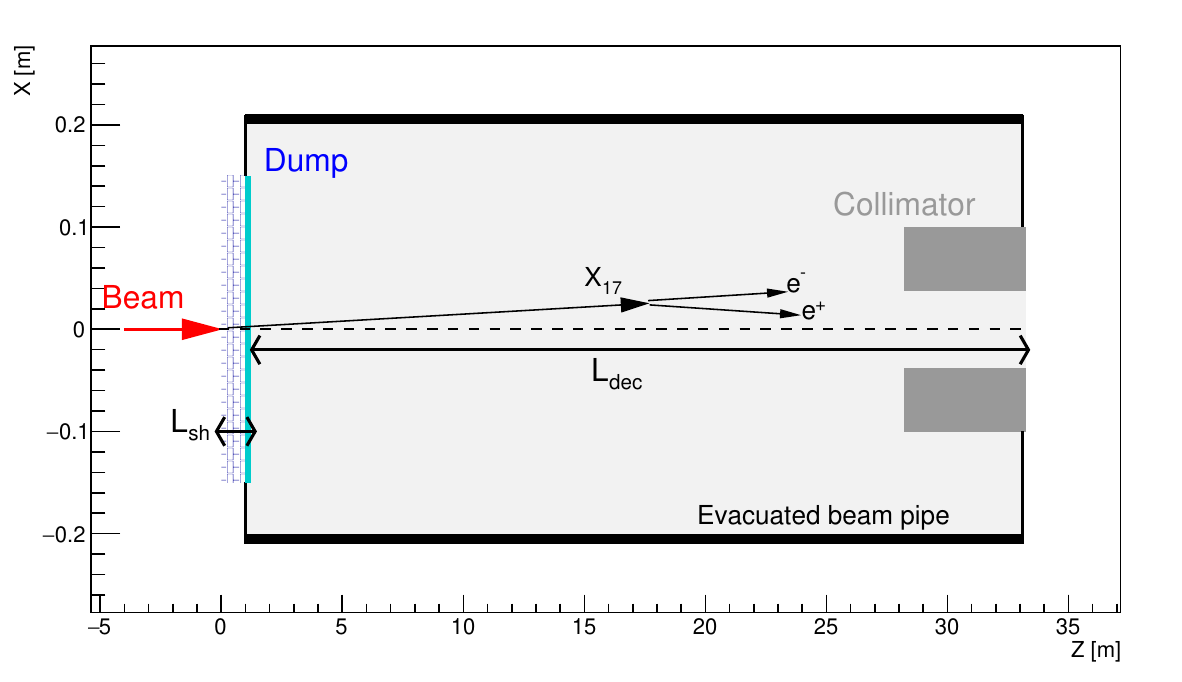}
    \caption{Schematic drawing of the E141 experimental setup. The cyan portion of the tungsten target corresponds to the additional 2-cm absorber installed during the second part of the run. The SLAC 8-GeV spectrometer detector, not shown in the diagram, was installed after the collimator, at 35 m distance from the target. For illustration, the production of a $X_{17}$ particle in the dump and the subsequent decay to an $e^+e^-$ pair in the downstream evacuated beam pipe is shown. For better visualization, the dump thickness has been magnified by a factor $\times 10$.}
    \label{fig:E141_setup}
\end{figure}

The SLAC E141 experiment~\cite{Riordan:1987aw} was originally designed to search for long-lived neutral pseudoscalar particles produced within a thick tungsten target by an impinging 9-GeV electron beam, having an intrinsic angular dispersion of about 0.2~mrad and a negligible momentum spread. Positrons from the in-flight decay of pseudoscalars emerging from the dump were detected by the so-called ``SLAC 8-GeV spectrometer''~\cite{Andivahis:1992zk}, aligned with the primary beam axis and located 35 m downstream of the dump - an evacuated beam pipe was installed between the detector and the dump to minimize multiple scattering and absorption of  positrons in air. The angular acceptance was defined by a 5-m long, 7.5-cm diameter collimator installed in the last section of the evacuated beam pipe, selecting particles with an angle of $\approx 1.1$~mrad with respect to the nominal beam direction. The detector assembly included a set of ten proportional wire chambers, allowing for momentum reconstruction with $\simeq 0.1\%$ resolution, a Cerenkov counter, and a lead-glass calorimeter for energy measurement and particle identification. The calorimeter was sensitive to positrons with energy $E_{p}$ within the range $4.5~\mathrm{\, GeV}<E_p<8.1\mathrm{\,GeV}$ ($0.5 \leq x_p \leq 0.9$, $x_p\equiv E_p/E_0$). A simplified diagram of the experimental setup is reported in Fig.~\ref{fig:E141_setup}.

During the E141 run, two different measurements were executed, each with a total accumulated charge of about $2\times10^{15}$ EOT - the systematic uncertainty on the overall normalization was $\approx 5\%$. In the first measurement, a $10.16$~cm target was employed, while in the second the dump length was increased to $12.16$~cm. Given the finite target length, totalizing 27.5 $X_0$ (32.2 $X_0$) in the first (second) configuration, for both setups a sizeable positron rate was measured, decreasing from $\approx2.4\times10^{-11}$/GeV/EOT for $x_{p}=0.45$ to $\approx1.1\times10^{-12}$/GeV/EOT for $x_{p}=0.9$ (from $\approx8.9\times10^{-13}$/GeV/EOT to $6.7\times10^{-14}$/GeV/EOT). This yield was interpreted as arising from the production of an energetic photon within the shower which penetrates the dump and pair-converts into an electron-positron pair in the very last portion of it. To maximize the experiment sensitivity, the original analysis combined the two results according to this interpretation, assuming the reduction in the measured rate with the $12.16$~cm dump to be entirely due to the increased photon absorption probability due to the enhanced target thickness. 
From this, E141 set competitive exclusion limits for the existence of axions in the mass range between 1 and 15 MeV and  lifetimes between $10^{-14}$~s and $10^{-10}$~s.

In 2012,  the original data reported by E141 was re-interpreted by Andreas and collaborators in the context of dark photon ($A^\prime$) searches at electron beam-dump experiments~\cite{Andreas:2012mt}. Dark photons are hypothetical $J^{P}=1^-$ massive particles predicted by many Standard Model extensions~\cite{Fabbrichesi:2020wbt} featuring a new $U(1)$ gauge symmetry associated to a new interaction in Nature. The theoretical description of dark photon interactions with SM states differs from that of Eq.~\ref{eq:theory}, since in the first scenario a universal coupling with SM charged leptons $\varepsilon eq$ is assumed, $q$ being the particle electric charge in elementary charge units. 
Nevertheless, the phenomenology of $X_{17}$ production via electromagnetic processes in an electron-beam thick-target experiment closely resembles that of the $A^\prime$ for the same mass and electron coupling values, and thus  results from the aforementioned work can be directly transported to the $X_{17}$ physics case. 

The E141 results were reported in terms of the positron energy spectrum, $dN_{p}/dE_p$, normalized to the number of electrons on target. The experimental uncertainties associated with this measurement related to the beam-charge normalization and to geometrical and acceptance effects were accounted for in the reported positron yield~\cite{Riordan:1987aw}. Using these data, Ref.~\cite{Andreas:2012mt} obtained an effective positron yield in the signal region $0.7\leq x_{p}\leq0.9$ of $N_{e^+}=(1126^{+1132}_{-1126})$. The uncertainty on this quantity therefore already incorporates the experimental systematic contributions entering the original E141 measurement. From this result, Ref.\cite{Andreas:2012mt} derived upper limits on the number of signal events of $N_s^{95}=3419$ at $95\%$ CL ($N_s^{90}=2965$ at $90\%$ CL), in which these uncertainties are thus already accounted for. Through the use of Eq.~\ref{eq:totYield}, the limits on the signal yield were then translated into an exclusion contour in the $\varepsilon_e$-vs-$m_X$ parameter space, excluding $5.1\times10^{-5}\lesssim\varepsilon_e\lesssim1.7\times10^{-4}$ for $m_X=16.88$~MeV.

The work of Ref.~\cite{Andreas:2012mt} established the benchmark framework for deriving competitive exclusion limits on dark photons - and thus $X_{17}$ - existence from past beam-dump electron-beam experiments data, in continuity with the first phenomenological arguments discussed earlier in Ref.~\cite{Bjorken:2009mm}. Nevertheless, the analysis introduced some simplifications and approximations that could non-negligibly affect the final result. Specifically, we observe that in calculating the total signal yield the IWW formula was adopted, instead of the more rigorous ETL result. Also, multiple scattering and beam divergence effects were ignored in the evaluation of the experimental acceptance, and the broadening of $\theta_{prod}$ due to primary electron energy degradation in the target was not accounted for. Finally, the $X_{17}$ production vertex longitudinal coordinate within the target was artificially fixed to $z=0$, ignoring the corresponding distribution. In the following, we present a new re-analysis of the E141 dataset under the $X_{17}$ hypothesis, including all these effects 
and quantifying the impact of each on the final result. Our finding is that no upper limit can be set for $X_{17}$, i.e. no $\varepsilon_e$ value exists such that $N_s \ge N_s^{UP}$, if $m_X \gtrsim 16$~MeV.

Our approach is based on a full Monte Carlo simulation of the E141 experimental setup, modelling the electromagnetic shower development in the target and computing the $X_{17}$ production through the use of the ETL total cross section. To optimize the computation pipeline, a two-stage approach was followed. First, the differential $X_{17}$ production yield $\frac{dN}{dE_X d\cos(\theta) d\bm{r}}$ was computed, where $\bm{r}$ is the $X_{17}$ production vertex within the dump. In the second step, the overall signal yield was calculated, accounting for the $\varepsilon_e$-dependent detector acceptance. This procedure was repeated for different values of $m_X$ to explore the whole parameters space of the $X_{17}$ model. 

\subsection{\label{sec:X17prod}$X_{17}$ production in the E141 target}

\begin{figure}[t]
    \centering
    \includegraphics[width=0.5\textwidth]{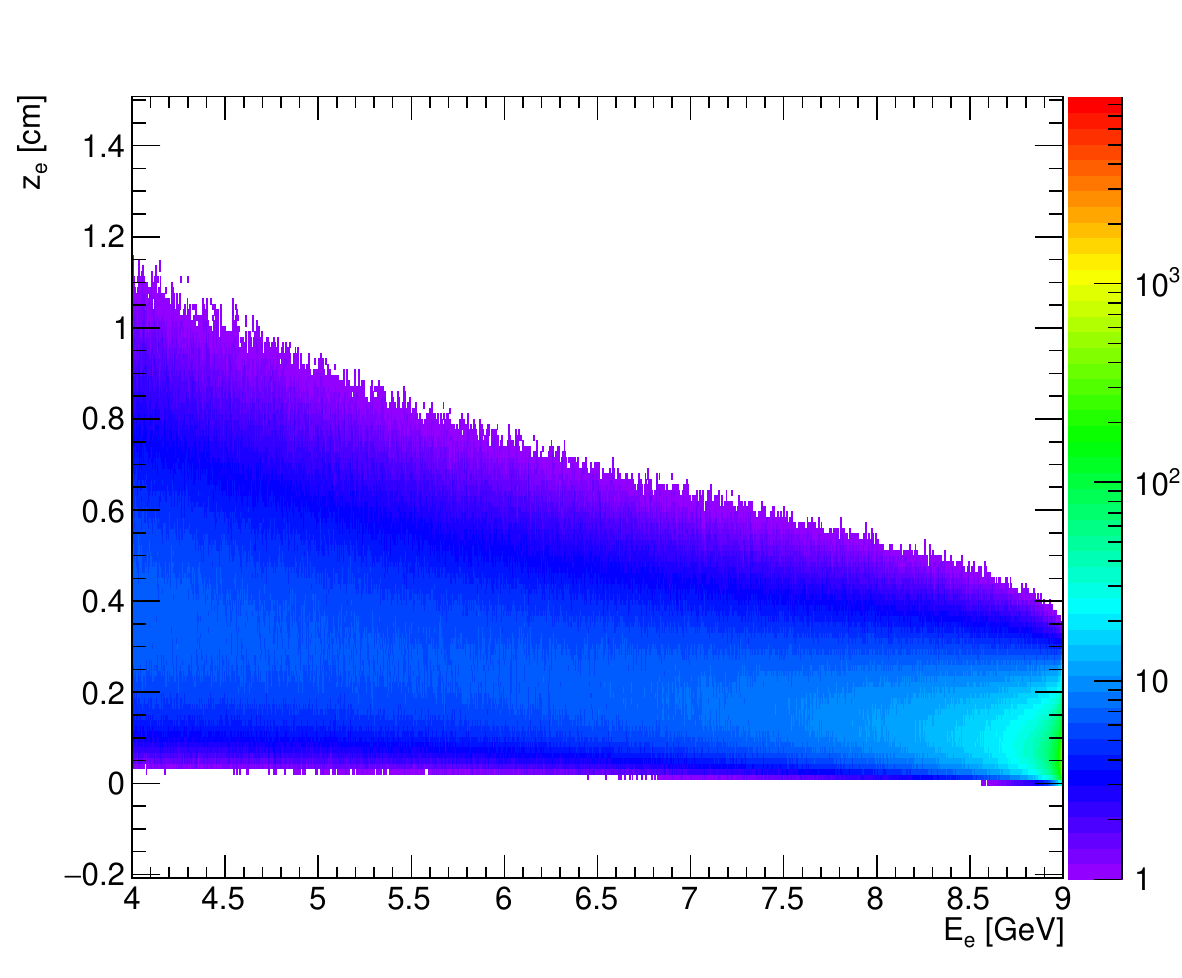}
    \caption{Bidimensional distribution of electrons longitudinal coordinate $z$ vs energy $E_e$ in the E141 beam-dump.}
    \label{fig:EvsZ}
\end{figure}

We computed the $X_{17}$ total production yield inside the E141 target through the \texttt{MADDUMP} plugin~\cite{Buonocore:2018xjk} for the \texttt{MadGraph5 aMC$@$NLO} (MG5) program~\cite{Alwall:2011uj}. \texttt{MADDUMP} extends MG5 functionalities to simulate microscopic interactions in a thick-target electron-beam setup, accounting for the the energy and angular distribution of the initial-state particle. As input, the code requires a numerical parameterization of the differential electrons track-length $T(E_e,\cos \theta_e,\phi_e)$ in the target. This was obtained from a Geant4-based simulation of the 9 GeV $e^-$ beam interaction with the E141 target.  According to the description reported in Ref.~\cite{Riordan:1987aw} the beam angular divergence was parametrized as flat-$\cos(\theta)$ distribution, $\theta$ being the angle with respect to the nominal beam axis, for $0\le \theta \le0.2~\mathrm{mrad}$. The output of the simulation included the total yield, computed for fixed value of the electron coupling $\varepsilon_e=10^{-4}$, and a list of $X_{17}$ events through which the detector acceptance was computed, as discussed in the next section. Since the production vertex distribution is not described in \texttt{MADDUMP}, it was effectively accounted for a posteriori, constructing the electrons $E_e$-vs-$z$ bidimensional distribution within the shower by marginalizing the track-length over the electron angle, and then randomly assigning to each MC event a $z-$coordinate value randomly extracted from the 1-D distribution corresponding to that event initial-state electron energy. The bidimensional distribution is shown in Fig.~\ref{fig:EvsZ} - even for $E_e \simeq E_0$, $z$ values up to 0.5~cm are observed, corresponding to $\approx 5\%$ of the E141 target length. Given the exponential dependence of the longitudinal acceptance on the effective target thickness, this shows that the $z-$coordinate distribution can't be neglected in the analysis, especially for the larger $\varepsilon_e$ values explored. As an example, if one considers a 5-GeV $X_{17}$, with  $\varepsilon_e = 2 \times 10^{-4}$, a production vertex of $z=0.5$ results in an increased probability of the decay occurring outside the dump of $\approx 15 \%$, compared to a production vertex occurring at $z=0$.  

To cross-check the results, the calculation was repeated  employing the DMG4 package~\cite{Celentano:2021cna,Oberhauser:2024ozf} to model the vector boson production by radiative emission from either the primary electron or a secondary one within the electromagnetic shower. An ad-hoc biasing scheme was introduced in the simulation to optimize the computation time. First, given the E141 spectrometer energy range, the total $X_{17}$ production cross section was artificially set to zero for electron energy lower than $E_{0}^{thr}=4$~GeV. Second, for $E>E_{0}^{thr}$ it was scaled by a $m_X$-dependent biasing factor $\beta$, tuned to result in $X_{17}$ production for about $1\%$ of the simulated events. Finally, a custom Geant4 \texttt{G4StackingAction} class was implemented to allow to selectively run the full shower simulation only for events in which $X_{17}$ production above the energy threshold actually occurred (see e.g. Ref.~\cite{NA64:2025rib}, Appendix B, for a complete description). Similarly to the previous case, the simulation provides as output the overall normalization and a list of unbiased MC events.

\begin{figure}
    \centering
    \includegraphics[width=0.5\textwidth]{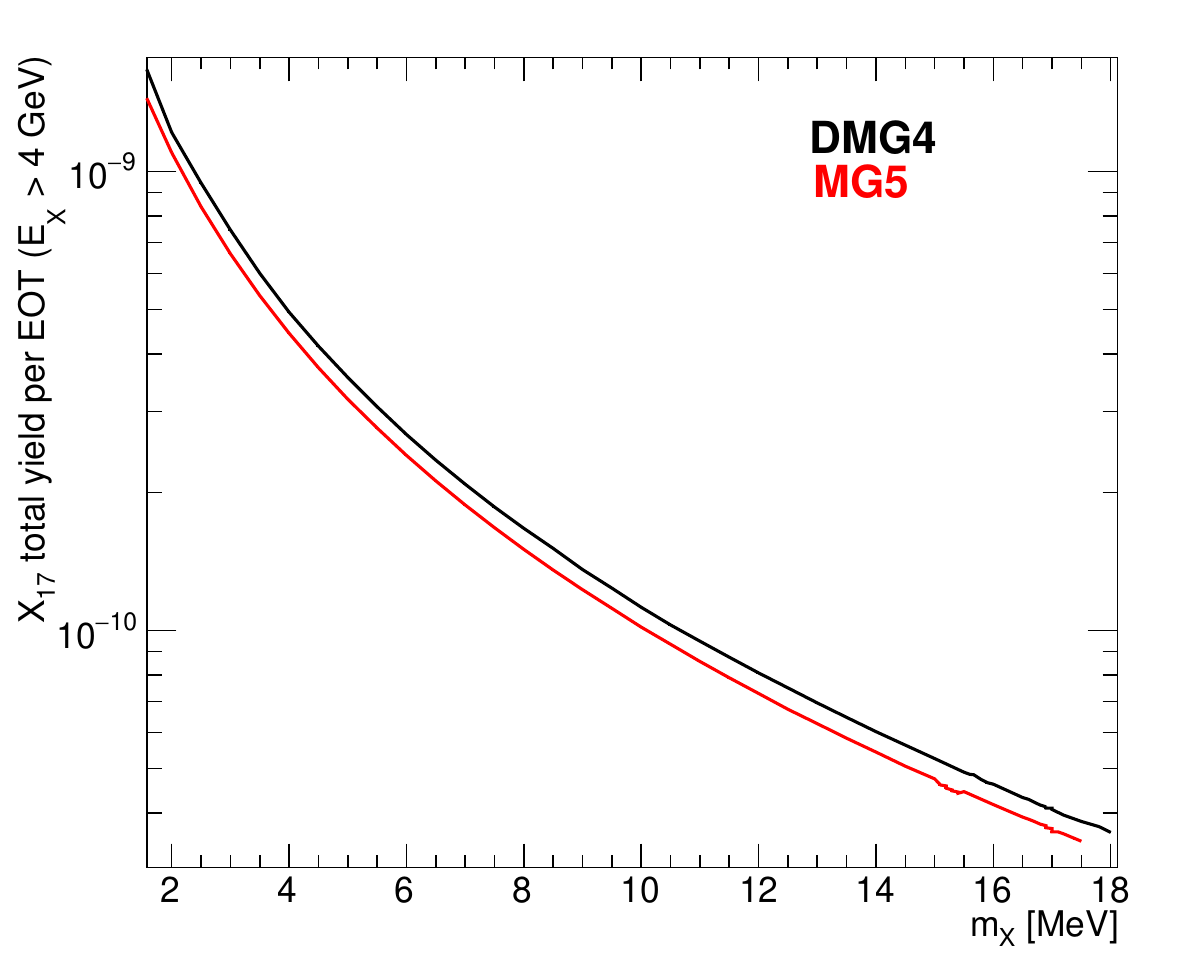}
    \caption{Total $X_{17}$ production yield per EOT in the E141 dump for $E_{X}>4$~GeV, as a function of $m_{X}$, for $\varepsilon_e=10^{-4}$. The black (red) curve has been obtained from a calculation based on DMG4 (\texttt{MADDUMP}). See text for further details.}
    \label{fig:yield}
\end{figure}

The total $X_{17}$ production yield per EOT for the emission of a $X_{17}$ particle with $E_{X}>4$~GeV is reported in Fig.~\ref{fig:yield} as a function of $m_X$, for $\varepsilon_e=1$, comparing the DMG4 result (black) with the \texttt{MADDUMP} one (red). We observe that the two codes produce very similar results, demonstrating the robustness of our approach. We also mention that the \texttt{MADDUMP} yield is systematically lower than the DMG4 one, the difference being $\approx 10\%$ over the whole $m_X$ range relevant for this work. 
In this analysis, we decided to focus on the \texttt{MADDUMP} result, cross-checking it with the DMG4 one. This choice is motivated by the fact that \texttt{MADDUMP} samples the final state kinematics from the fully-differential ETL cross section, including spin-density matrix effects in the $X_{17}$ decay process, while DMG4 generates the vector boson energy distribution employing the IWW approximation, while normalizing the total production cross section to the ETL result, and assumes an isotropic decay mechanism. Nevertheless, we conservatively accounted for the discrepancy between the two results as a systematic uncertainty of the yield calculation. We anticipate that, owing to the strong dependency of the event yield on $\varepsilon_e$, this effect has a limited impact for the final result.

\subsection{Signal yield in the E141 detector}

The signal efficiency was evaluated using the Monte Carlo sample produced as described in the previous section. For each generated event and for a given value of $\varepsilon_e$, the boosted $X_{17}$ decay length $\lambda_D$ was computed and the decay position sampled according to the corresponding exponential decay law. Only events decaying downstream the target inside the evacuated beam pipe were retained. The kinematics of the $X_{17}\to e^+e^-$ decay, including the angular correlations arising from the $X_{17}$ polarization, were taken directly from the \texttt{MADDUMP} event sample generated from the full matrix element. An event was considered accepted if the positron trajectory intersected the E141 collimator aperture, with the corresponding energy falling within the $x_p$ signal range. The detector acceptance, which depends on both the $X_{17}$ mass and coupling, was then computed as
\begin{equation}
    \eta(m_X,\varepsilon_e)=\frac{N_{\rm in}}{N_{\rm all}}\,,
\end{equation}
where $N_{\rm in}$ is the number of accepted events and $N_{\rm all}$ is the total number of generated events. From this, the total signal yield $N_s$ was computed.

\begin{figure}
    \centering
    \includegraphics[width=0.5\textwidth]{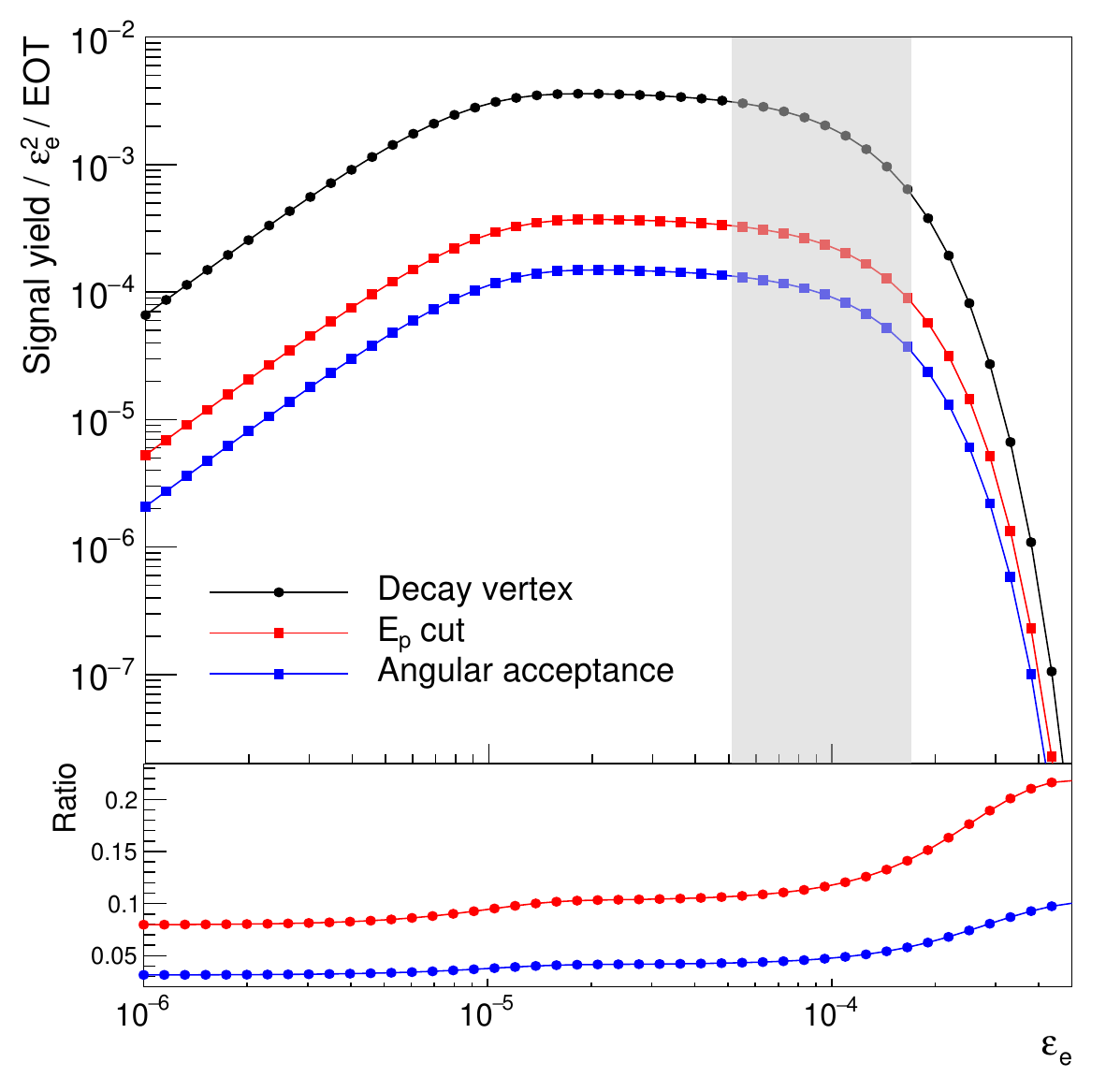}
    \caption{\textbf{Top panel:} experimental E141 yield per EOT, normalized to $\varepsilon^2$, for the reaction $e^-N\rightarrow e^-NX_{17}$ followed by the in-flight decay $X_{17}\rightarrow e^+e^-$ reported as a function of $\varepsilon_e$ for $m_X=16.88$~MeV. The black curve has been obtained from a simplified MC simulation based directly on Eq.~\ref{eq:totYield}, not including any angular effect and integrating over the whole $x_p$ range, while the red one includes the $0.7\leq x_p \leq 0.9$ energy selection. The blue curve further includes the effect of the detector angular acceptance. The gray area highlights the $\varepsilon_e$ exclusion interval from Ref.~\cite{Andreas:2012mt}. \textbf{Bottom panel:} fraction of the decay-vertex yield retained after applying the positron-energy selection (red) and, additionally, the detector angular acceptance (blue).}
    \label{fig:acceptance}
\end{figure}

For illustration, the signal yield per EOT as a function of $\varepsilon_e$, normalized by $\varepsilon_e^2$, is shown in Fig.~\ref{fig:acceptance}, for $m_X=16.88$~MeV, evaluated on events with $E_X>4$~GeV. In the top panel, the black curve corresponds to the result obtained from Eq.~\ref{eq:totYield}, without accounting for any angular effect, and considering the whole positron energy range, while the red one includes the selection $0.7 \leq x_p \leq 0.9$; finally, the blue curve additionally accounts for the angular effects. The gray area highlights the current $X_{17}$ exclusion interval from Ref.~\cite{Andreas:2012mt}. As expected, in the low-coupling region ($\varepsilon_e \lesssim 10^{-5}$) a quadratic dependency is observed, owing to the almost-uniform decay vertex distribution, resulting to a decay probability within the acceptance region $P \propto {L_{dec}}{\lambda_D} \propto \varepsilon_e^2$. In the opposite large-coupling range, instead, the normalized yield is dominated by the first exponential factor in Eq.~\ref{eq:totYield}, parametrizing the $X_{17}$ probability to decay downstream the target. Finally, this observable shows a smoother trend in the intermediate region, as a consequence of the approximate compensation between two exponential factors in Eq.~\ref{eq:totYield}. 

Interestingly the fraction of the decay-vertex yield retained after applying the
positron-energy selection, as well as that obtained including also the angular acceptance (bottom panel) shows a large suppression approximately constant over nearly two orders of magnitude in coupling, $10^{-6}\lesssim\varepsilon_e\lesssim10^{-4}$. The increased ratio for $\varepsilon_e \approx 10^{-3}$, instead, originates from kinematic correlations, since in this large coupling regime, only events with $E_X \approx E_0$, with strongly-boosted decay length, contribute to the signal yield - for these, the kinematics of $X_{17}$ production and subsequent decay is strongly peaked forward, making angular effects less pronounced. Despite the nearly constant ratio, the observed reduction in the detection acceptance has markedly different effects on the lower- and upper-coupling exclusion boundaries. In particular, for the latter the strong steepness of the $\eta(\varepsilon_e)$ curve causes the limit to be determined primarily by the position of the acceptance turnover occurring when $\lambda_D < L_{sh}$, rather than by its overall normalization. This can be qualitatively justified considering that, for coupling values $\varepsilon_e \approx 10^{-4}$, the acceptance can be roughly approximated as a step-function $\eta(\varepsilon_e) \approx \eta_0 \theta(\varepsilon_e^0-\varepsilon_e)$, with $\varepsilon_e^0 \approx 1.5\times 10^{-4}$ - in this case, the experiment is always insensitive to $\varepsilon_e > \varepsilon_e^0$, no matter the value of $\eta_0$. For the lower coupling boundary, instead, the observed reduction in the experimental acceptance translates to $\mathcal{O}(\mathrm{few})$ reduction of the limit, owing to the $\varepsilon_e^2$ dependency of the production yield.

\subsection{Results}

To obtain the new $X_{17}$ exclusion contour in the $\varepsilon$-vs-$m_x$ plane we started from the 95$\%$ upper limit for the absolute signal yield in the E141 experiment computed in Ref.~\cite{Andreas:2012mt}, $N_s^{95}=3419$. For each hypothesized $m_X$ value, considered in increasing order from $m_X=1.1$~MeV to $m_X=20$~MeV, the condition $N_s<N_s^{95}$ was translated to an exclusion interval for $\varepsilon_e$ by inverting the yield equation (Eq.~\ref{eq:totYield}), properly accounting for all experimental effects discussed in the previous sections. We adopted a non-constant grid for the mass scan, with steps ranging from 500 keV at low mass to 10 keV close to the edge of the exclusion limit contour. Owing to the non-linear dependency of $N_s$ over $\varepsilon_e$, a numerical iterative approach was adopted. First, $N_s$ was computed for a grid of 100 points $P_i$ for $\varepsilon_e$, logarithmically spaced from $\varepsilon_{DOWN}$ to $\varepsilon_{UP}$ - for better accuracy, a finer 200-points grid was used for mass values larger than 15 MeV. Exploiting the expected shape of the exclusion contour, these were respectively set to the lower and upper boundary of the exclusion interval obtained for previously-tested $m_X$ value - for the lower $m_X=1.1$~MeV point, these were respectively initialized to $\varepsilon_{DOWN}=10^{-5}$ and $\varepsilon_{UP}=10^{-2}$. Then, to find the lower boundary $\varepsilon_e^{min}$, the two consecutive grid points $P_i$, $P_{i+1}$ satisfying $N_s^{i}<N_s^{95}<N^{i+1}_s$ were located. The interval $(\varepsilon_e^{i},\varepsilon_e^{i+1})$ was iteratively refined until the point $P^{min}$ was found for which the corresponding yield $N^{min}$ agreed with $N_s^{95}$ within $0.1\%$. A similar procedure was adopted for the upper boundary. 

\begin{figure}
    \centering
    \includegraphics[width=0.5\textwidth]{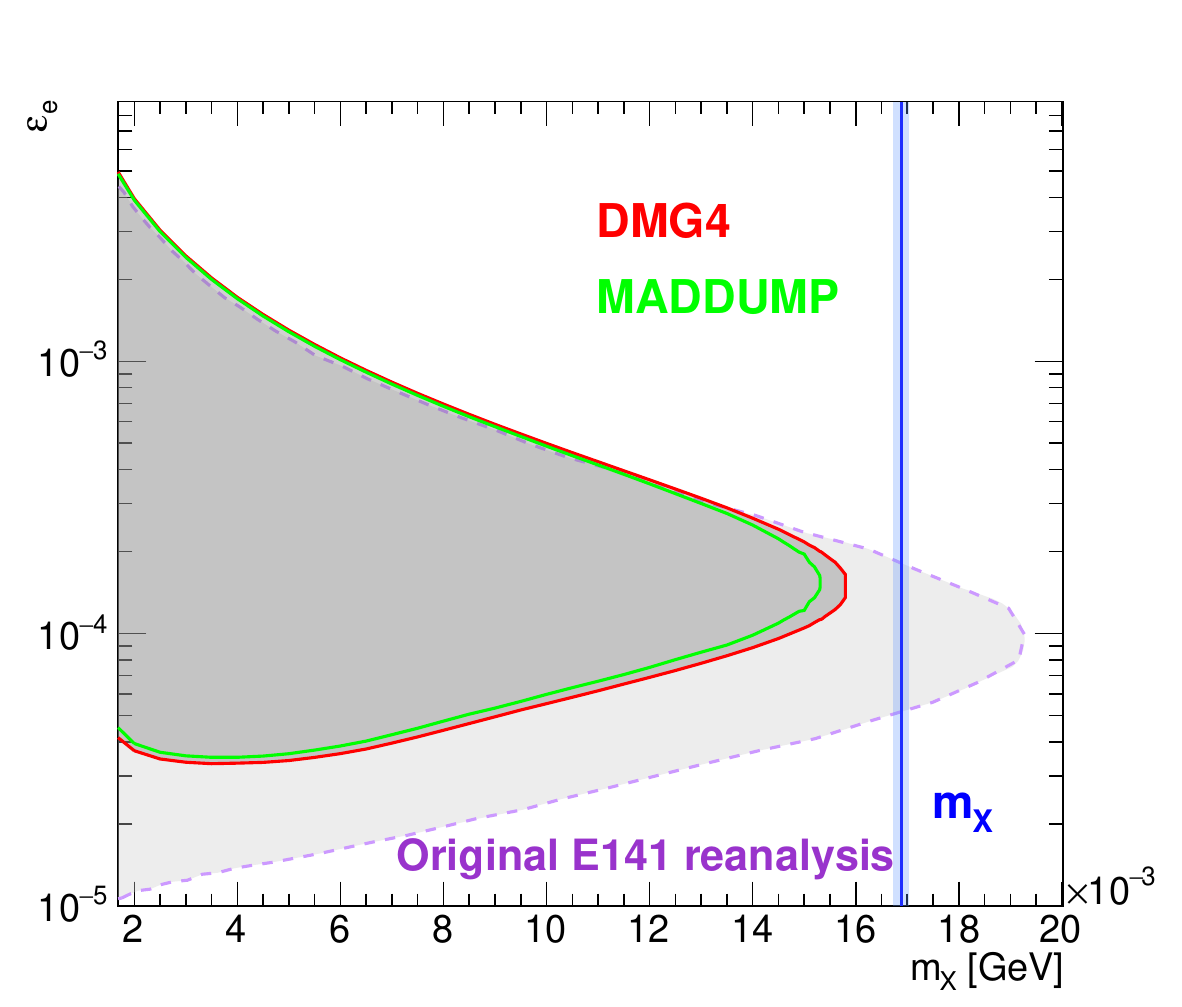}
    \caption{Updated 95$\%$~CL exclusion limits from E141 for the existence of the $X_{17}$ particle, as a function of $m_X$ and $\varepsilon_e$. The red (green) curve refers to the result obtained using the DMG4 (\texttt{MADDUMP}) software to compute the $X_{17}$ yield within the E141 beam dump. The results are compared with the original 
    limit from Ref.~\cite{Andreas:2012mt} (magenta curve). The blue band shows the estimated $X_{17}$ mass $m_X=(16.88\pm0.05)$~MeV from Ref.~\cite{Arias-Aragon:2025wdt} - for better visualization, the corresponding uncertainty band, in light blue, has been magnified by a factor $\times 3$.}
    \label{fig:result95}
\end{figure}

The new limit is shown in Fig.~\ref{fig:result95}, where the 95$\%$ CL exclusion contour obtained in this analysis is compared with the original result from Ref.~\cite{Andreas:2012mt}. The two effects reducing the event yield that were not previously accounted for, namely the use of the ETL parameterization for the production cross section and the inclusion of all angular effects in the detection acceptance, significantly impact the result in the lower $\varepsilon_e$ branch, leaving almost unchanged the upper one, as anticipated in the previous Section. From this result, we conclude that the original E141 dataset, reanalysed in the context of a search for visibly-decaying $X_{17}$ particle, does not allow any exclusion limit on the $\varepsilon_e$ coupling to be set for mass values greater than 15.31~MeV (15.81~MeV) at 95$\%$ CL (90$\%$ CL). 

As discussed in Sec.~\ref{sec:E141}, the systematic uncertainties associated with the original E141 measurement, related to the beam-charge normalization and to geometrical and acceptance effects, are already included in the signal-yield upper limits derived in Ref.~\cite{Andreas:2012mt} and adopted in the present analysis. They are therefore not considered as additional systematic uncertainties on the exclusion contour. The remaining systematic uncertainty relevant to our re-analysis is associated with the modelling of the $X_{17}$ production yield in the target. To estimate this, we compared the exclusion limit derived using the \texttt{MADDUMP} software to compute the $X_{17}$ yield with that obtained using the DMG4 package. Since DMG4 by default does not consider spin-density matrix effects in the $X_{17} \rightarrow e^+e^-$ decay, for proper comparison with the \texttt{MADDUMP} results we made the assumption of a fully-transversely polarized $X_{17}$, resulting to a positron angular distribution in the decay frame proportional to $1+cos^2(\theta_e)$, $\theta_e$ being the $e^+$ emission angle measured with respect to the $X_{17}$ direction in the laboratory frame. The two results are in good agreement over almost the entire mass range, reflecting the small $(\lesssim 10\%)$ difference between the predicted $X_{17}$ yield from the two simulations. A noticeable discrepancy is visible only near the upper end of the sensitive mass range, since there the yield variation directly translates to a $\simeq 3.1 \%$ difference in the mass end-point of the exclusion contour, at both $90\%$ CL and $95\%$ CL. Even accounting for this additional systematic uncertainty on the final result, still the new exclusion contour does not reach the $m_X=(16.88 \pm 0.05)$ mass interval. Uncertainties in the apparatus geometrical acceptance, as clearly stated in~\cite{Riordan:1987aw}, are already included in the E141 normalized positron yield systematic error, and thus accounted for in this work. Nevertheless, we performed a supplementary robustness check by recalculating the 95\% CL exclusion region while varying the relevant geometrical parameters, in order to assess the stability of our results. For what concerns the target thickness, the original work describes it as a 12.16-cm thick block: we thus assumed an uncertainty of 0.01 cm in the length, as deduced from the last significant digit. For $X_{17}$ parameters close to the edge of the exclusion contour, the particle mean free path in the laboratory is larger than 4.5 cm, considering the positron energy threshold adopted in the original experiment. This value is 450 times larger than the above uncertainty, that can thus be safely neglected. For the distance between the target and the collimator and the collimator radius, since no precise uncertainty estimate is reported in~\cite{Riordan:1987aw}, we assumed a best-guess of $\pm 10$ cm and $\pm 2$ mm, respectively. Our study showed that the limit is only mildly dependent on the distance, while a variation of the collimator dimension, instead, has a more pronounced effect - if the radius is increased by 2 mm, the exclusion limit boundary extends from 15.32 MeV to 15.50 MeV. The $X_{17}$ predicted mass value of ($16.88 \pm 0.05$) MeV is still well outside the newly-derived E141 exclusion interval.

\begin{figure}[t]
    \centering
    \includegraphics[width=0.5\textwidth]{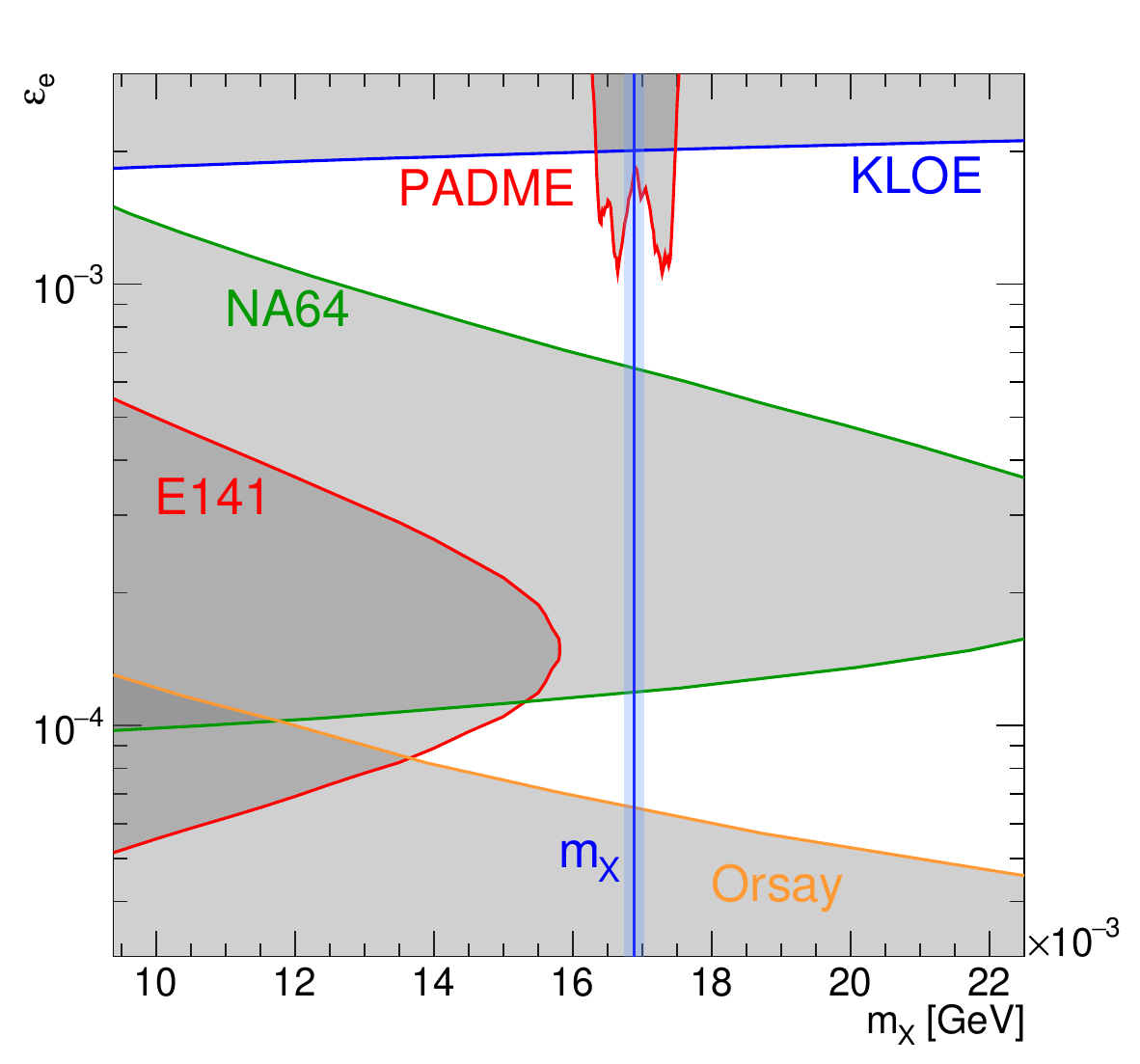}
    \caption{Summary of current $X_{17}$ exclusion limits at $90\%$~CL from E141 (current analysis), NA64~\cite{NA64:2019auh}, KLOE~\cite{Anastasi:2015qla}, PADME~\cite{PADME:2025dla}, and the Orsay beam-dump experiment~\cite{Davier:1989wz}. The blue line indicates the region of the parameter space compatible with the recent preliminary measurement of the $X_{17}$ lifetime by ATOMKI~\cite{x17longlived,x17longlived1} - the line width reflects the estimated $X_{17}$ mass $m_X=(16.88 \pm 0.05)$~MeV from Ref.~\cite{Arias-Aragon:2025wdt}, magnified by a factor $\times 3$ for better visualization.}
    \label{fig:result90}
\end{figure}

As anticipated in the Introduction, this opens the possibility for the existence of a long-lived $X_{17}$ particle, with $\varepsilon_e \lesssim 10^{-4}$, compatible with the 
range identified by the recent preliminary measurements of the corresponding lifetime~\cite{x17longlived}. To show this, we provide in Fig.~\ref{fig:result90} an updated summary of current $90\%$ CL exclusion limits for this particle from experiments directly sensitive to the electron coupling $\varepsilon_e$, comparing results from NA64~\cite{NA64:2019auh}, KLOE~\cite{Anastasi:2015qla}, PADME~\cite{PADME:2025dla}, as well as from the Orsay beam-dump experiment~\cite{Davier:1989wz}, re-interpreted within the dark-photon physics case in Ref.~\cite{Andreas:2012mt}. For the latter, the measured null event yield within the signal region was translated by Andreas \textit{et al.} to a $95\%$~CL limit for a visibly-decaying dark photon by setting $N_s^{UP}=3.0$. In principle, the $90\%$~CL scenario here considered, corresponding to $N_s^{UP}=2.3$, would call for a new re-analysis of these past results. This goes beyond the scope of this work. Here, we adopted a simple correction, considering that, for the large-coupling boundary, the $N_s$ dependency on $\varepsilon_e$ is dominated by the exponential factor giving the probability to decay downstream the beam-dump length $L_{sh}$. To compute the corresponding decay length, we assumed at first order $E_X \approx E_0$ - for the Orsay experiment, $E_0=1.6$~GeV and $L_{sh}=1$~m, scaling the limit accordingly. The overall correction was found to be well below $1\%$ for all mass points. As seen from the Figure, the newly-derived E141 exclusion limit discussed in this work opens a new window in the $X_{17}$ parameters space, for $6.5\times10^{-5}\lesssim \varepsilon_e \lesssim 1.1\times 10^{-4}$.

\section{Conclusions}
In this paper we perform a recalculation of the E141 exclusion limits in the vector boson interpretation of the $X_{17}$ anomaly. 
While the analysis presented in Ref.~\cite{Andreas:2012mt} provided a valuable  assessment of the sensitivity of E141 to a generic light vector boson, the precise $X_{17}$ mass determination emerging from the combined observations of ATOMKI, MEG II, and PADME call for a more refined treatment of the E141 results, commensurate with the present level of experimental precision. This is not least because the preferred $X_{17}$ mass, $(16.88\pm0.05)$ MeV, lies close to the edge of the excluded region, where seemingly minor acceptance effects neglected in previous reinterpretations can have a significant impact on the derived exclusion limits, as already pointed out in Ref.~\cite{Nardi:2018cxi}.

Building upon the framework of Ref.~\cite{Andreas:2012mt}, we implemented the refinements required for a precise determination of the exclusion limits on the $X_{17}$ coupling. In particular, the angular effects associated with the beam dispersion, electromagnetic shower development in the thick-target, $X_{17}$ production, and propagation of the decay products were evaluated via a detailed simulation of the involved processes, accounting for the precise geometry of the E141 experimental setup. These effects lead to a non-negligible reduction in the signal acceptance, largely independent of the $\varepsilon_e$ value, significantly weakening the E141 sensitivity in the 14 -- 18~MeV mass region of the parameter space.  As a consequence, the revised exclusion limits no longer extend to the $X_{17}$ mass of 16.88~MeV, leaving a sizeable region of parameter space for the vector-boson interpretation of the anomaly, bounded by the NA64 and Orsay constraints. Interestingly, the remaining allowed interval, $6.5\times10^{-5}\lesssim \varepsilon_e \lesssim 1.1\times10^{-4}$, is compatible with the recent preliminary measurement of the $X_{17}$ lifetime, setting the constraint $\varepsilon_e \gtrsim 2.6\times10^{-5}$. We further observe that, for these coupling values, the $X_{17}$ decay length for a 100~GeV/c $e^-$ beam-dump experiment is of the order of a few meters, resulting to an invisible signature for a missing-energy setup such as NA64-$e$~\cite{NA64:2023wbi}. The $X_{17}$ parameters space not any longer constrained by E141 could thus be explored in the near future by NA64 through the invisible-mode dataset accumulated so far by the experiment.

In conclusion, this work shows that reinterpreting legacy experiments to constrain specific BSM scenarios may require a substantially higher level of accuracy than that needed for an initial sensitivity estimate. 
At the same time, our results reinforce the case for new dedicated experiments capable of directly probing the remaining viable parameter space of the $X_{17}$ vector-boson hypothesis.

\begin{acknowledgments}

This result is part of a project that has received funding from the European Research Council (ERC) under the European Union’s Horizon 2020 research and innovation programme, Grant agreement No. 947715 (POKER). This work was supported by the Istituto Nazionale di Fisica Nucleare (INFN). A.C. warmly thanks the organizers of the ``X17 What if?" workshop (INFN-LNF, 19-20 March 2026), where the initial idea developed in this work arose, for their kind hospitality. 
\end{acknowledgments}

\vspace{.5cm}

\appendix
\section{\label{appendix:MM}Explicit formulas for the $X_{17}$ radiative cross section at exact-tree-level}

In this appendix, we compute the exact tree-level differential cross section for radiative $X_{17}$ emission in the process $e^- N \rightarrow e^- N X_{17}$. We emphasize that our results are in agreement with those reported in Ref.~\cite{Gninenko:2017yus}. We present the calculation here for completeness and to correct a few errors that we identified in the expressions given in that reference (eq. A14).

We will denote with $p^\mu=(E,\bm{p})$ (${p^\prime}^\mu=(E^\prime,\bm{p}^\prime)$) the four-momentum of the impinging (scattered) electron, with $P_i^\mu=(M,0)$ ($P_f^\mu=(E_f,\bm{p}_f)$) that of the target (recoil) nucleus, and with $k^\mu=(E_k,\bm{k})$ that of the emitted particle. The masses of the electron, nucleus, and $X_{17}$ are, respectively, $m_e$, $M$, $m_X$. The four-momentum transferred to the nucleus is $q^\mu=P^\mu_f-P^\mu_i$\footnote{We adopt a convention with an opposite sign with respect to Refs.~\cite{Liu:2017htz,Gninenko:2017yus}.}. Adopting the mostly-minus Lorentz metrics, the differential cross section is given by the expression:
\begin{widetext}
\begin{equation}\label{eq:dsigma1}
    d\sigma = \frac{\left |\overline{\overline{\mathcal{M}}} \right |^2}{4M p \left(2\pi\right)^5}\frac{d\bm{k}d\bm{p^\prime}d\bm{P_f}}{8E_kE^\prime E_f}\delta(E+M-E_k-E^\prime-E_f)\delta(\bm{p}+\bm{P_i}-\bm{k}-\bm{p^\prime}-\bm{P_f}) \; \; ,
\end{equation}
\end{widetext}
where $\left|\overline{\overline{\mathcal{M}}} \right |^2$ is the matrix element squared for the process, averaged and summed over the helicities of initial and final state particles, respectively. Without loss of generality, we schematize the nucleus as a spin-0 target, whose interaction with the exchanged photon is described by a vertex containing a momentum-dependent form factor $F(q^2)$, $-ieF(q^2)(P_i^\mu+P_f^\mu)\equiv ieF(q^2)\mathcal{P}^\mu$. We adopt the same functional form for $F(q^2)$ introduced in Ref.~\cite{Bjorken:2009mm}, Eq.~A18 and A19.

We integrate over the momentum-conservation $\delta$ function by fixing $\bm{p^\prime}=\bm{p}-\bm{q}-\bm{k}$, where $\bm{q}=\bm{P_f}$ in the laboratory frame with the target nucleus is at rest, to get:

\begin{equation}\label{eq:dsigma2}
    d\sigma = \frac{\left |\overline{\overline{\mathcal{M}}} \right |^2}{1024 \,M \,p \,\pi^5}\frac{d\bm{k}d\bm{q}}{E_k \bar{E}^\prime E_f}\delta(E+M-E_k-\bar{E}^\prime-E_f) \; \; ,
\end{equation}
where $\bar{E}^\prime = \sqrt{(\bm{p}-\bm{q}-\bm{k})^2+m_e^2}$. This expression can be further worked out writing explicitly $d\bm{q}=q^2dkq d\cos(\theta_{q})d\phi_q$, where $\theta_q$ is the angle that the $\bm{q}$ vector forms with the $\bm{p}-\bm{k}$ one and $\phi_q$ parametrizes the residual azimuthal rotational degree of freedom. The remaining energy-conservation $\delta$ function can be exploited to integrate-out $\cos(\theta_{q})$. By writing the energy-conservation constraint as:
\begin{align}\label{eq:ene}
    E+M-&E_k-E_f=\sqrt{(\bm{p}-\bm{q}-\bm{k})^2+m^2_e}= \nonumber
    \\
    &=\sqrt{(\bm{p}-\bm{k})^2+|\bm{q}|^2-2|\bm{q}|\left|\bm{p}-\bm{k}\right|\cos(\theta_q)+m_e^2} \; \;,
\end{align}
we get:
\begin{equation}
\delta(E+M-E_k-\tilde{E^\prime}-E_f) = \delta(\cos(\theta_{q}) - \cos(\theta_q^*))\,\frac{\widetilde{E}^\prime}{|\bm{q}|\left|\bm{p}-\bm{k}\right|} \; \; ,
\end{equation}
where $\cos(\theta_q^*)$ is the value of $\cos(\theta_q)$ given by Eq.~\ref{eq:ene}, and ${\widetilde{E}}^\prime \equiv \bar{E}^\prime(\cos(\theta_q^*))$. We thus obtain the expression:
\begin{equation}\label{eq:dsigma3}
  d\sigma = \frac{\left |\overline{\overline{\mathcal{M}}} \right |^2}{1024 \,M \,|\bm{p}| \,\pi^5}
  \frac{q dq d\phi_qd\bm{k}}{ E_k E_f \left|\bm{p}-\bm{k} \right|} = \frac{\left |\overline{\overline{\mathcal{M}}} \right |^2}{1024 \,M \,
  |\bm{p}| \,\pi^5}
  \frac{dE_f d\phi_qd\bm{k}}{E_k \left|\bm{p}-\bm{k} \right|} \; \; .
\end{equation}
To proceed further and perform the integration over $\phi_q$, we 
take the explicit expression for $\left|\overline{\overline{\mathcal{M}}} \right |^2$ reported in Ref.~\cite{Gninenko:2017yus}, Eq. 12, that we cross-checked independently in this work:
\begin{equation}\label{eq:M}
\left|\overline{\overline{\mathcal{M}}}\right|^2 = e^6 \varepsilon_e^2 \frac{\left|F(t)\right|^2}{t^2}|A|^2 \; \; , 
\end{equation}
where
{\small
\begin{align}\label{eq:A}
|A|^2=&
\frac{2}{\s^2 \u^2}
\Bigl(
2 m_e^2 \left(\P^2 t (\s+\u)^2-4((\P \cdot p) \u +(\P \cdot p') \s)^2\right)+\nonumber \\ 
+&m_{A'}^2 \left(\P^2 t (\s-\u)^2-4((\P \cdot p) \u +(\P \cdot p') \s)^2\right)+\nonumber \\
+&\s \u \left(\P^2 ((\s +t)^2+(\u+t)^2) -4 t\left[( \P \cdot p)^2 +( \P \cdot p')^2 \right]\right)\Bigr).
\end{align}
}
The Mandelstam variables and relevant Lorentz products are:
\begin{align}
\s &=(p'^\mu+ k^\mu)^2-m_e^2 = (p^\mu-q^\mu)^2-m^2_e, \nonumber \\
\u & =(p^\mu - k^\mu)^2-m_e^2  \nonumber \\
t & = -({P_i}^\mu-{P_f}^\mu)^2 = -q_\mu q^\mu\nonumber \\
\P^2 &= 4 M^2+t, \nonumber \\
\P \cdot p & = 2ME  -(\s+t)/2 \nonumber \\
\P \cdot p' & = 2ME^\prime+(\u+t)/2.
\end{align}
In these expressions, the dependency on $\phi_q$ is contained solely in $\tilde{s}$.
To make this explicit, we need a specific expression for the $\bm{q}$ vector. We introduce two unitary vectors $\hat{u}_a$ and $\hat{u}_b$ such that $\hat{u}_a$ lies in the $\bm{p}, \bm{k}$ plane and is perpendicular to $\bm{p}-\bm{k}$, and $\hat{u}_b= \hat{u}_{\bm{p}-\bm{k}} \times \hat{u}_a$, so that  
\begin{equation}
\bm{q}=|\bm{q}|\left(\cos(\theta^*_q)\hat{u}_{\bm{p}-\bm{k}}+\sin(\theta_q^*)\cos(\phi_q)\hat{u}_a+\sin(\theta_q^*)\sin(\phi_q)\hat{u}_b\right) \; \; ,
\end{equation}
to get
\begin{equation}
\tilde{s}=-t-2p_\mu q^\mu=-t-2E(E_f-M)+2\bm{p}\cdot\bm{q} 
\end{equation}
with the scalar product reading
\begin{equation}
\bm{p}\cdot\bm{q}=|\bm{p}|\,|\bm{q}|\left(\cos(\theta_q^*)\cos(\theta_{pk})+\sin(\theta_q^*)\sin(\theta_{pk})\cos(\phi_q)\right) \; ,
\end{equation}
where $\theta_{pk}$ is the angle between $\bm{p}$ and $\bm{p}-\bm{k}$.
The relevant integrals to be performed are:
\begin{widetext}
\begin{align}
I_{-2}=\frac{1}{2\pi}\int_0^{2\pi}\frac{d\phi_k}{s^2}=&
\frac{\left|-t-2E(E_f-M)+2|\bm{p}|\, |\bm{q}|\cos(\theta_q)\cos(\theta_{pk})\right|}
{\left[\left(-t-2E(E_f-M)+2|\bm{p}|\, |\bm{q}|\cos(\theta_q)\cos(\theta_{pk})\right)^2
-4|\bm{p}|^2\, |\bm{q}|^2\sin^2(\theta_q)\sin^2(\theta_{pk})\right]^{3/2}}
\nonumber\\
I_{-1}=\frac{1}{2\pi}\int_0^{2\pi}\frac{d\phi_k}{s}=&
\frac{\mathrm{sgn}\left\{-t-2E(E_f-M)+2|\bm{p}|\, |\bm{q}|\cos(\theta_q)\cos(\theta_{pk})\right\}}
{\left[\left(-t-2E(E_f-M)+2|\bm{p}|\, |\bm{q}|\cos(\theta_q)\cos(\theta_{pk})\right)^2
-4|\bm{p}|^2\, |\bm{q}|^2\sin^2(\theta_q)\sin^2(\theta_{pk})\right]^{1/2}}
\nonumber\\
I_{0}=\frac{1}{2\pi}\int_0^{2\pi}d\phi_k=&\,1 \nonumber\\
I_{1}=\frac{1}{2\pi}\int_0^{2\pi}s\,d\phi_k=&
-t-2E(E_f-M)+2|\bm{p}|\, |\bm{q}|\cos(\theta_q)\cos(\theta_{pk}) \; ,
\end{align}
where $\mathrm{sgn}(x)\equiv |x|/x$. Combining everything together, we find:
\begin{equation}
    \frac{1}{2\pi}\int_0^{2\pi} d\phi_k\left|A\right|^2 = I_1 C_1 + I_0C_0 + I_{-1} C_{-1} +I_{-2} C_{-2} \;  \; ,
    \end{equation}
with
\begin{align}
 C_1 = & \,8 \frac{M^2}{\u} 
\\
C_{0} = &\frac{-8M\left(m_X^2+2m_e^2\right)\left(4{E^\prime}^2M+2{E^\prime}\,t-Mt\right)+16M\left(E+M\right)t\u}{\u^2}
\\
C_{-1}=&\,\frac{8}{\u}\Bigg(
-8EE_fM^2\left(2m_e^2+m_X^2\right)
-2M\Big[
2\left(E-E^\prime\right)m_e^2
+2\left(E^2+{E^\prime}^2-m_e^2\right)M
+\left(E-E^\prime+M\right)m_X^2
\Big]t+ \nonumber\\
&\qquad\qquad
+2\left(m_e^2+M\left(E-E^\prime+M\right)\right)t^2
+2M\left(M-E_f\right)t\,\u
+M^2\u^2
\Bigg)
\\
C_{-2}=&-8M\left(2m_e^2+m_X^2\right)\left(4E^2M-2Et-Mt\right)\;.
\end{align}
From these expressions we finally get the exact tree-level cross section $\frac{d\sigma}{dE_k d\cos(\theta_k)}$ for the $e^- N \rightarrow e^- N X_{17}$ reaction:
\begin{equation}\label{eq:dsigmadxdq}
\frac{d\sigma}{dE_kd\cos(\theta_k)dt}=\frac{\alpha^3\varepsilon_e^2}{8 M^2|\bm{p}|}\frac{|\bm{k}|}{\left|\bm{p}-\bm{k}\right|}\frac{\left| F(t)\right|^2}{t^2}{\left[I_1C_1+I_0C_0+I_{-1}C_{-1}+I_{-2}C_{-2}\right]} \; \;.
\end{equation}
The kinematic limits on the momentum transferred squared $t$ follow from Eq.~\ref{eq:ene}, imposing the mathematical constraint $-1\leq\cos(\theta_q)\leq1$. After some algebra we get:
\begin{align}
t_{\rm min,max} = &\,
\frac{
2M^2\left(|\bm{p}|^2+|\bm{k}|^2-2|\bm{p}
|\,|\bm{k}|\cos(\theta_k)\right)
-M\left(M+E-E_k\right)\left(2EE_k-m_X^2-2|\bm{p}
|\,|\bm{k}|\cos(\theta_k)\right)
}{
\left(M+E-E_k\right)^2-\left(|\bm{p}|^2+|\bm{k}|^2-2|\bm{p}
|\,|\bm{k}|\cos(\theta_k)\right)
}
\nonumber\\
&\mp
\frac{
2M|\bm{p}|\,|\bm{k}|\sqrt{|\bm{p}|^2+|\bm{k}|^2-2|\bm{p}|\,|\bm{k}|\cos(\theta_k)}
}{
\left(M+E-E_k\right)^2-\left(|\bm{p}|^2+|\bm{k}|^2-2|\bm{p}
|\,|\bm{k}|\cos(\theta_k)\right)
}\times
\nonumber\\
&\times
\sqrt{
\left(\cos(\theta_k)-\alpha_-)\right)
\left(\cos(\theta_k)-\alpha_+)\right)
}\, \; ,
\end{align}
with
\begin{align}
\alpha_\pm=
\frac{
2E(E_k-M)-m_X^2+2M(E_k\pm m_e)
}
{2|\bm{p}|\,|\bm{k}|}\;\;.
\end{align} 
Finally, the allowed interval for $\cos(\theta_k)$ is given by $\left(\alpha_+\le\cos(\theta_k)\le1 \right)\cap \left(\cos(\theta_k)\ge-1\right)$.
\end{widetext}

\bibliographystyle{apsrev4-2}
\bibliography{bibliographyNA64_inspiresFormat.bib,bibliographyNA64exp_inspiresFormat.bib,bibliographyOther_inspiresFormat.bib} 

\end{document}